\documentclass[]{llncs}

\usepackage{amsmath}
\usepackage{xspace}
\usepackage{graphicx}
\usepackage{mymacros}

\title{Iterators, Recursors and Interaction Nets}

\author{Ian Mackie\inst{1} \and Jorge Sousa Pinto\inst{2} \and Miguel
  Vila\c{c}a\inst{2}\\
\textsf{mackie@lix.polytechnique.fr,\{jsp,jmvilaca\}@di.uminho.pt}}

\institute{LIX, CNRS UMR 7161, \'Ecole Polytechnique, 91128 Palaiseau Cedex, 
France\and Departamento de Inform\'atica / CCTC, Universidade do Minho, Braga, Portugal}

\begin{document}

\maketitle

\begin{abstract}
  We propose a method for encoding iterators (and recursion operators
  in general) using interaction nets (INs). There are two main applications
  for this: the method can be used to obtain a visual notation for
  functional programs; and it can be
  used to extend the existing translations of the $\lambda$-calculus
  into INs %
to languages with recursive types.
\end{abstract}

\section{Introduction}

The use of visual notations for functional programs has long
been an active research topic, whose main goal is to have a notation
that can be used (i) to define functional programs visually, and (ii)
to animate their execution. %

In this paper we propose a graphical system for functional
programming, based on token-passing INs~\cite{DBLP:conf/tlca/Sinot05}. The system
offers an adequate solution for classic problems of visual notations,
including the treatment of higher-order functions, pattern-matching,
and recursion (based on the use of recursion operators). The system
implements a call-by-name semantics, with a straightforward
correspondence between functional programs and graphical objects.

Most approaches to visual programming simply propose a notation for
programs. Program evaluation is animated by representing visually the
intermediate programs that result from executing reduction steps on
the initial program, using the operational semantics of the underlying
functional language. Our approach is different in that we use a
graph-rewriting formalism with its own semantics.

The advantages of using INs for visual programming are:

\begin{itemize}

\item Both programs and data are represented in the same %
  graphical formalism.

\item Programs can be animated without leaving the interaction
  formalism. %

\item Pattern-matching for external constructors is in-built. 

\item Recursive definitions are expressed very naturally as
  interaction rules involving agents %
that are
  reintroduced on the right-hand side. %

\end{itemize}

But the interaction net
formalism does not offer a satisfactory semantic interpretation for
the behaviour of functional symbols. Moreover, many interaction net systems
can be defined that do not have a functional reading.
What is missing is a clear correspondence between functional
definitions and interaction systems.%
In this paper
we establish a correspondence between agents with ``obviously
functional'' interaction rules %
and
functions defined with recursion operators.

\section{The Token-passing Encoding of the \lbdc}
\label{sec:tokpass}

\noindent
The \emph{token-passing} encodings~\cite{DBLP:conf/tlca/Sinot05} use
an interaction system where two different symbols exist for
application: one is the syntactic symbol $@$ introduced by the
translation; the corresponding agents have their principal ports
facing the root of the term and will be depicted by triangles. A
second symbol $\widehat{@}$ exists that will be used for computation;
to simplify the figures, the corresponding agents will be depicted by
circles equally labelled with $@$. Their principal ports face the net
that represents the applied function, to make possible interaction
with $\lambda$ agents.

The translation $\Ttp{\cdot}$ encodes terms in the system $(\Sigtp,
\Rtp)$ where $\Sigtp = \{\da, @, \widehat{@}, \lambda, c, \eraser,
\delta \}$. %
It generates nets containing no active pairs.%
The special symbol $\da$ is used as an evaluation \emph{token}: an
agent $\da$ traverses the net, transforming occurrences of $@$ into
$\widehat{@}$. %
The evaluation rules
involving $\da$ can be tailored to a specific evaluation strategy.
For call-by-name, $\Rtp$ consists of the following rules %
which comprises evaluation rules involving
$\da$ and a computation rule involving $@$ and $\lambda$. Management
(copying and erasing) rules are omitted here. %
  \begin{center}
  \includegraphics[width=.65\textwidth]{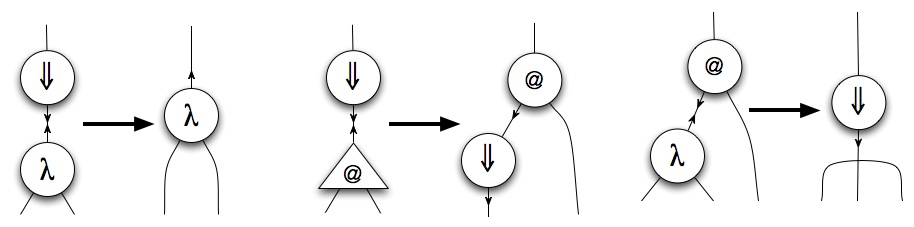}
\end{center}

To start the reduction %
a
$\da$ symbol must be connected to the root port of the term. Let
$\token N$ denote such net; %
 then the following correctness result holds: $
\evals{t}{z}\; \mbox{iff}\; \token \Ttp{t} \lra^* \Ttp{z}$, where the
evaluation relation $\evals{\cdot}{\cdot}$ is defined by the standard
evaluation rules for call-by-name.

The language used in this paper is the simply-typed \lbdc extended
with natural numbers, booleans, lists, and iterators for these
types. \ourlang is defined by the following syntax for %
terms
($x$, $y$ range over a set of variables):
\vspace{-0.2cm}
\small
\[
\begin{array}{rcl}
  t,u,v & ::= & x \mid \abs{x}{t} \mid \ap{t}{u} \mid \true \mid \false \mid \iterbool{t}{u}{v}
   \mid  0 \mid \suc{t} \mid \iternat{\abs{x}{t}}{u}{v}  \\
& \mid &  \nil \mid \cons{t}{u}  \mid \iterlist{\abs{xy}{t}}{u}{v}
\end{array}
\]
\normalsize

\section{A Token-passing Encoding of \ourlang}
\label{sec:second-approach}

We extend to \ourlang the token-passing \cbnm translation of the \lbdc
into the interaction system $(\Sigtp, \Rtp)$. %
The novelty of
this encoding is not the token-passing aspect, %
 but rather the approach to
recursion.

We first consider data structures.  %
In a %
token-passing implementation, there will be an interaction rule
between the token agent and each %
constructor symbol that will
stop evaluation. %
For \ourlang we define the system $(\Sigol, \Rol)$ where $\Sigol$
consists of the symbols $\true$, $\false$, 0 and $\nil$ with arity 0;
$\mathsf{suc}$ with arity 1; and $\mathsf{cons}$ with arity 2. %
Each recursive program will be encoded in an interaction system
specifically generated for it. This is a major novelty of our
approach. The interaction system for the \lbdc will not be extended by
introducing a fixed set of symbols; instead a new symbol will be
introduced for \emph{each occurrence of a recursion operator}, with an
interaction rule for each different constructor, %
so a dedicated interaction system $(\Sigt^0, \Rt^0)$ is generated for
each term $t$.
This system is constructed by a recursive function $(\Sigt^0, \Rt^0) =
\Sn{t}$, defined as: %
\vspace{-0,6cm}
  \begin{center}
  \small
  \[
  \begin{array}[]{l}
    \Sn{x} \doteq \Sn{\true} \doteq \Sn{\false} \doteq \Sn{0} \doteq
    \Sn{\nil} \doteq (\emptyset, \emptyset)\\

    \Sn{\abs{x}{t}} \doteq \Sn{\suc{t}} \doteq \Sn{t}\\

    \Sn{\ap{t}{u}} \doteq \Sn{\cons{t}{u}} \doteq \Sn{t} \cup \Sn{u} \\

    \Sn{\iterbool{V}{F}{b}} \doteq
    (\{\itboolag{V}{F}, \widehat{\itboolag{V}{F}} \} \cup \Sigma,
    R_{\itboolag{V}{F}} \cup R), \\ 
    \quad \parbox{.85\textwidth}{ where 
      $(\Sigma, R) = \Sn{b} \cup \Sn{V} \cup \Sn{F}$.
	}  \\ 

    \Sn{\iternat{\abs{x}{S}}{Z}{n}} \doteq
    (\{\itnatag{S}{Z}, \widehat{\itnatag{S}{Z}}\} \cup
    \Sigma,R_{\itnatag{S}{Z}} \cup R)\\   
    \quad \parbox{.85\textwidth}{where 
      $
      (\Sigma, R) = \Sn{n} \cup \Sn{S} \cup \Sn{Z}
      $ 
}\\ 

    \Sn{\iterlist{\abs{xy}{C}}{N}{l}} \doteq
    (\{\itlstag{C}{N}, \widehat{\itlstag{C}{N}} \} \cup \Sigma, R_{\itlstag{C}{N}} \cup R)\\  
    \quad \parbox{.85\textwidth}{where 
      $
      (\Sigma, R) = \Sn{l} \cup \Sn{C} \cup \Sn{N}
      $ 
}\\  
  \end{array}
  \]
  \normalsize
\end{center}

$R_{\itlstag{C}{N}}$ consists of the following interaction rules (the others are similar).

\centerline
  { \includegraphics[width=1.6 in]{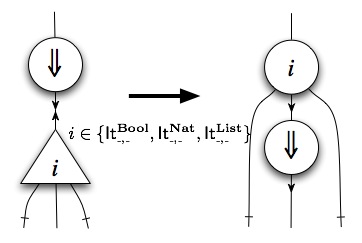}}%

\centerline{\label{fig:list-iterator-rules}
\includegraphics[width=.4\textwidth]{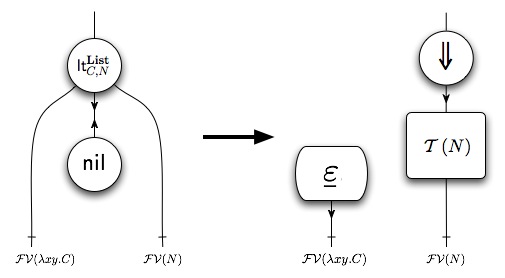}\qquad
\includegraphics[width=.4\textwidth]{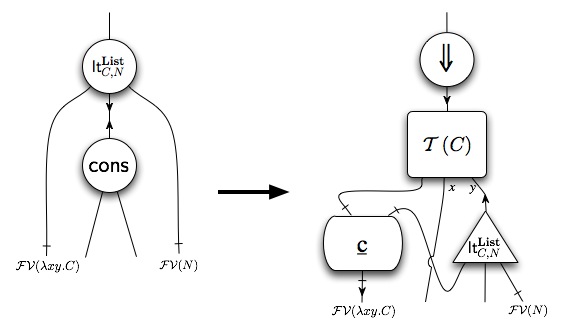}}%

Iterator symbols are introduced in pairs
$(\mathsf{It}_{\ldots}^{\ldots},
\widehat{\mathsf{It}_{\ldots}^{\ldots}})$ where the first symbol is
used for syntactic agents and the second for computation agents
(similarly to $@, \widehat{@}$). %

A \ourlang program $t$ will be translated into a net defined in the
system $ (\Sigt, \Rt) = (\Sigtp \cup \Sigol \cup \Sigt^0, \Rtp \cup
\Rol \cup \Rt^0) $ where $(\Sigtp, \Rtp)$ was defined in
Section~\ref{sec:tokpass}.

Given a \ourlang program $t$, where $t$ is $ \iterbool{V}{F}{b}$, 
$\iternat{\abs{x}{S}}{Z}{n}$ or $\iterlist{\abs{xy}{C}}{N}{l}$, then 
the net $\T{t}$ is given as follows.

\begin{center}
\includegraphics[width=1.2in]{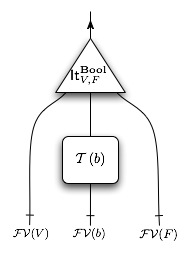}
\includegraphics[width=1.2in]{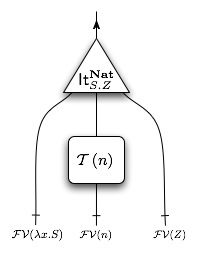}%
\includegraphics[width=1.2in]{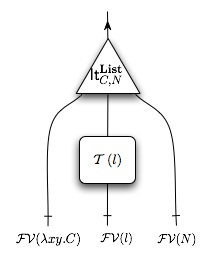}
\end{center}

In token-passing implementations, all terms
are translated as syntax trees. Syntactic
iterator agents $i$ are turned into their computation counterparts
$\widehat{i}$ by token agents. %
A first key aspect of our approach is that the interaction rules of
the (computation) iterator agents internalise the iterator's
parameters. For instance the net
$\T{\iterlist{\abs{xy}{C}}{N}{\cons{h}{t}}}$ reduces to
$\token \T{\substd{C}{x}{h}{y}{{\iterlist{\abs{xy}{C}}{N}{t}}}}$. %

A second key aspect is that each such new symbol will have auxiliary
ports in a one-to-one correspondence with the free variables in the
iterator term. %
We end the section with a correctness result.  The proofs can be found
in~\cite{vilacaURL}.

\begin{proposition}
({\bf Correctness})
If $t$ is a closed \ourlang term and $z$ a canonical form, then:
$\evals{t}{z} \quad \Longleftrightarrow \quad \token \T{t} \lra^*\T{z}$.
\end{proposition}

\section{Conclusions and Future Work}
\label{sec:conclusions}

We have presented an approach to encoding in INs
functional programs defined with recursion operators, and given the
full details of the application of this approach to the token-passing
implementation of a call-by-name language, which results in a very
convenient visual notation for this language.  The approach can be
easily extended to richer sets of recursive types and other recursion
operators %
and also to new strategies.
The novel characteristics of the encoding are (i) the interaction
system is generated dynamically from the program, and (ii) the
internalisation of some of the parameters of the recursion operator in
the interaction rules. %
With respect to previous work on encoding recursion in interaction
nets, fixpoint operators have been studied elsewhere for interaction
net implementations~\cite{AspertiA:optifp,MackieIC:geoim}, and we have
shown elsewhere how a binding recursion operator (as in PCF) can be
implemented in the token-passing setting~\cite{PintoJS:tokpnfl}.

A prototype system for visual functional programming has been
developed, integrated in the tool
\textsf{INblobs}~\cite{PintoJS:toopin,VilacaM:inblobs} for interaction
net programming.  The tool consists of an evaluator for interaction
nets together with a visual editor and a compiler module that
translates programs into nets. The latter module allows users to type
in a functional program, visualize it, and then follow its evaluation
visually step by step. The current compiler module %
automatically
generates call-by-name %
or call-by-value
systems. Additionally, a visual editing mode is available that allows
users to construct nets corresponding to functional programs. In the
current implementation %
there is no way to convert visual
programs back to textual ones.

The token-passing translation is not however
very representative of most work in this area, which has concentrated
on designing \emph{efficient} translations; %
\cite{GonthierG:geoolr,MackieIC:yalyal,MackieIC:efflei} are some samples.
Let $\T{\cdot}$ be one such translation.  Typically $\T{\ap{t}{u}}$ is
constructed from $\T{t}$ and $\T{u}$ by introducing an application
symbol $@$ with its principal port connected to the root port of
$\T{t}$. Our treatment of iterators can be adapted to this setting by
removing the evaluator tokens and introducing the iterator agents with
the principal port immediately facing the argument.
When the iterated function is a closed term, a correctness result can
be easily established. %
Let $\abs{x}{S}$ be a closed term, then
$\T{\iternat{\abs{x}{S}}{Z}{0}} \lra \T{Z}$
and 
$\T{\iternat{\abs{x}{S}}{Z}{\suc{n}}} \lra
  \T{\subst{S}{x}{\iternat{\abs{x}{S}}{Z}{n}}}$

\bibliographystyle{abbrv}
\bibliography{bibfile}

\begin{thebibliography}{10}

\bibitem{vilacaURL}
J.~B. Almeida, I.~Mackie, J.~S. Pinto, and M.~Vila\c{c}a.
\newblock Encoding iterators in interaction nets.
\newblock Available from \verb+http://www.di.uminho.pt/~jmvilaca+.

\bibitem{PintoJS:tokpnfl}
J.~B. Almeida, J.~S. Pinto, and M.~Vila\c{c}a.
\newblock {T}oken-passing {N}ets for {F}unctional {L}anguages.
\newblock In J.~Giesl, editor, {\em Proceedings of the 7th International
  Workshop on Reduction Strategies in Rewriting and Programming (WRS'07)},
  volume 204 of {\em Electronic Notes in Theoretical Computer Science}, pages
  181--198, 2007.

\bibitem{PintoJS:toopin}
J.~B. Almeida, J.~S. Pinto, and M.~Vila\c{c}a.
\newblock A {T}ool for {P}rogramming with {I}nteraction {N}ets.
\newblock In {\em Proceedings of the The Eighth International Workshop on
  Rule-Based Programming (RULE'07)}, 2007.
\newblock To appear in Elsevier ENTCS.

\bibitem{AspertiA:optifp}
A.~Asperti and S.~Guerrini.
\newblock {\em The {O}ptimal {I}mplementation of {F}unctional {P}rogramming
  {L}anguages}, volume~45 of {\em Cambridge Tracts in Theoretical Computer
  Science}.
\newblock Cambridge University Press, 1998.

\bibitem{GonthierG:geoolr}
G.~Gonthier, M.~Abadi, and J.-J. L{\'e}vy.
\newblock The geometry of optimal lambda reduction.
\newblock In {\em Proceedings of the 19th {ACM} Symposium on Principles of
  Programming Languages (POPL'92)}, pages 15--26. ACM Press, Jan. 1992.

\bibitem{MackieIC:geoim}
I.~Mackie.
\newblock The geometry of interaction machine.
\newblock In {\em Proceedings of the 22nd ACM Symposium on Principles of
  Programming Languages (POPL'95)}, pages 198--208. ACM Press, January 1995.

\bibitem{MackieIC:yalyal}
I.~Mackie.
\newblock {YALE}: Yet another lambda evaluator based on interaction nets.
\newblock In {\em Proceedings of the 3rd International Conference on Functional
  Programming ({ICFP}'98)}, pages 117--128. ACM Press, 1998.

\bibitem{MackieIC:efflei}
I.~Mackie.
\newblock Efficient $\lambda$-evaluation with interaction nets.
\newblock In V.~van Oostrom, editor, {\em Proceedings of the 15th International
  Conference on Rewriting Techniques and Applications (RTA'04)}, volume 3091 of
  {\em Lecture Notes in Computer Science}, pages 155--169. Springer-Verlag,
  June 2004.

\bibitem{DBLP:conf/tlca/Sinot05}
F.-R. Sinot.
\newblock Call-by-name and call-by-value as token-passing interaction nets.
\newblock In P.~Urzyczyn, editor, {\em TLCA}, volume 3461 of {\em Lecture Notes
  in Computer Science}, pages 386--400. Springer, 2005.

\bibitem{VilacaM:inblobs}
M.~Vila\c{c}a.
\newblock Inblobs webpage.
\newblock http://haskell.di.uminho.pt/jmvilaca/INblobs/.

\end{thebibliography}

\end{document}